\documentclass[12pt]{iopart}
\usepackage{amssymb}
\usepackage{color}
\usepackage{mathrsfs}
\usepackage{graphics}
\usepackage{graphicx}
\usepackage{dcolumn}
\usepackage{bm,epsf}

\def\ni{\noindent}
\begin{document}
\title{Mergers of binary stars: The ultimate heavy-ion experience}

\author{Madappa Prakash 
and Sa{\v s}a Ratkovi{\' c} 
and James M. Lattimer }

\address{Department of Physics \& Astronomy, 
        State University of New York at Stony Brook,  
        Stony Brook, NY 11794-3800, USA}
\medskip
\address{E-mail: prakash@snare.physics.sunysb.edu and
  ratkovic@grad.physics.sunysb.edu and 
  lattimer@mail.astro.sunysb.edu}

\begin{abstract}
The mergers of black hole-neutron star binaries are calcuated using a
pseudo-general relativistic potential that incorporates ${\mathcal
O}(v^2/c^2)^3$ post-Newtonian corrections.  Both normal matter neutron
stars and self-bound strange quark matter stars are considered as
black hole partners.  As long as the neutron stars are not too massive
relative to the black hole mass, orbital decay terminates in stable
mass transfer rather than an actual merger.  For a normal neutron
star, mass transfer results in a widening of the orbit but the stable
transfer ends before the minimum neutron star mass is reached.  For a
strange star, mass transfer does not result in an appreciable
enlargement of the orbital separation, and the stable transfer
continues until the strange star essentially disappears.  These differences
might be observable through their respective gravitational wave signatures.
\end{abstract}
\maketitle

The closest analog of a high-energy heavy-ion collision in nature is
the gravitational wave-induced merger of two compact objects involving
at least one neutron star or strange quark matter star in a binary
system.  However, since such a collision would involve more than
$10^{57}$ particles, it is vastly more energetic and would represent the
ultimate heavy-ion experience.  Such events have been suggested to
occur frequently enough to account for some fractions of cosmological
gamma-ray bursts and of r-process heavy elements \cite{LSch}.

In this work, we contrast the evolution of binary star mergers for two
distinct cases: 

\ni (1) A black hole (BH) and a (mostly) normal matter neutron star with
a surface at which the pressure vanishes at vanishing baryon density.
The interior of the star, however, may contain any or a combination of
the many exotica such as hyperons, Bose (pion or kaon) condensate or
quark matter.

\ni (2) A BH and a self-bound star with a surface at which the
pressure vanishes at a large baryon density.  This case is exemplified
by a strange-quark matter (SQM) star~\cite{Alcock} with a bare quark
matter surface.

\begin{figure}[htb]
\begin{center}
\epsfxsize=\textwidth
\epsfbox{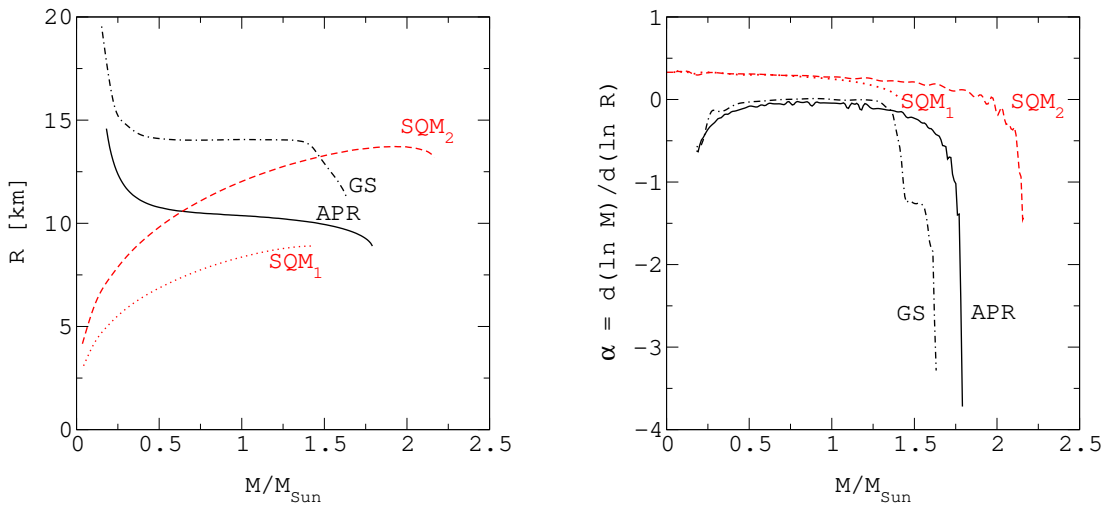}
\end{center}
\caption{\label{mnr}Radius versus mass (left panel) and its logarithmic
derivative for prototype EOS's. The EOS symbols are as in Ref. \cite{LP01}.}
\end{figure}

Prototypes of these cases are shown as mass-radius relations in
Fig.~\ref{mnr}. Quantitative variations from these generic behaviors
are caused by uncertainties in the strong interaction models (see the
compendium of results in Fig. 2 of Ref.~\cite{LP01}), but do not
lead to qualitative differences in gravitational mergers. The qualitative
differences between the two classes of mass-radius relations will, however,
produce significant changes in the outcomes of gravitational mergers.

A normal star and a self-bound star represent two quite different 
possibilities (right panel in Fig.~\ref{mnr}) for the quantity
\begin{eqnarray}
\alpha \equiv  \frac {d\ln R}{d\ln M} 
\left\{ \begin{array}{ll} 
\leq 0 & \mbox{{\rm for~a~normal~neutron~star~(NS)}} \\
\geq 0 & \mbox{{\rm for~a~self-bound~SQM~star}} 
\end{array} 
\right. \nonumber
\label{lderiv}
\end{eqnarray}
for small to intermediate masses, where $M$ and $R$ are the star's
mass and radius, respectively. For low mass self-bound
stars, $R \propto M^{1/3}$ so that $\alpha \cong 1/3$; only for masses
close to the maximum mass does $\alpha$ turn negative.
Note that $\alpha$ is intimately connected with the dense matter
equation of state (EOS), since there exists a one-to-one
correspondence between $R(M)$ and $P(n_B)$, where $P$ is the pressure
and $n_B$ is the baryon density.  Gravitational mergers in which a
compact star loses its mass (either to a companion star or to an
accretion disk) during their evolution comprise a rare example in
which the $R-M$ (or equivalently, the $P-n_B$) relation of a single
star is sampled.

Our objective here is to explore the astrophysical consequences of the
distinctive behaviors of the $R-M$ relation as they affect mergers
with a black hole (see also Refs. \cite{Lee01,Prakash03,Shibata03}).
In general, a gravitational merger begins with two widely-separated
objects with a mutual orbit decaying via gravitational radiation
reaction.  When the separation becomes small, the less massive
component can exceed its Roche limit and begin to lose mass to the
more massive star.  Typically, this can occur if the mass ratio
$q=M_1/M_2$ of the two stars ($M_1$ is the neutron star mass) is
somewhat less than unity.  During mass transfer, the radius of the
neutron star quickly readjusts to its new mass.  If the radius
increases as fast as the Roche limiting radius, mass transfer to the
BH will be stable, and the inspiral will be halted and reversed due to
angular momentum conservation.  During stable mass transfer, the
stellar radius and its Roche lobe remain coincident and mass transfer
can continue until the star's mass becomes very small.

However, the loss of angular momentum and energy due to gravitational
radiation reaction, as well as the extent of the Roche lobe, depend
upon the gravitational potential.  Until now, few merger simulations
have included general relativistic corrections to the potential for $q
\neq 1$.  Reference \cite{Blanchet02} has evaluated the effective
gravitational potential up to order $(v^2/c^2)^3$ in post-Newtonian
corrections.  Utilizing this potential, we have corrected the
gravitational radiation reaction and the effective Roche lobe radius.

\begin{figure}[htb]
\begin{center}
\epsfxsize=\textwidth \epsfbox{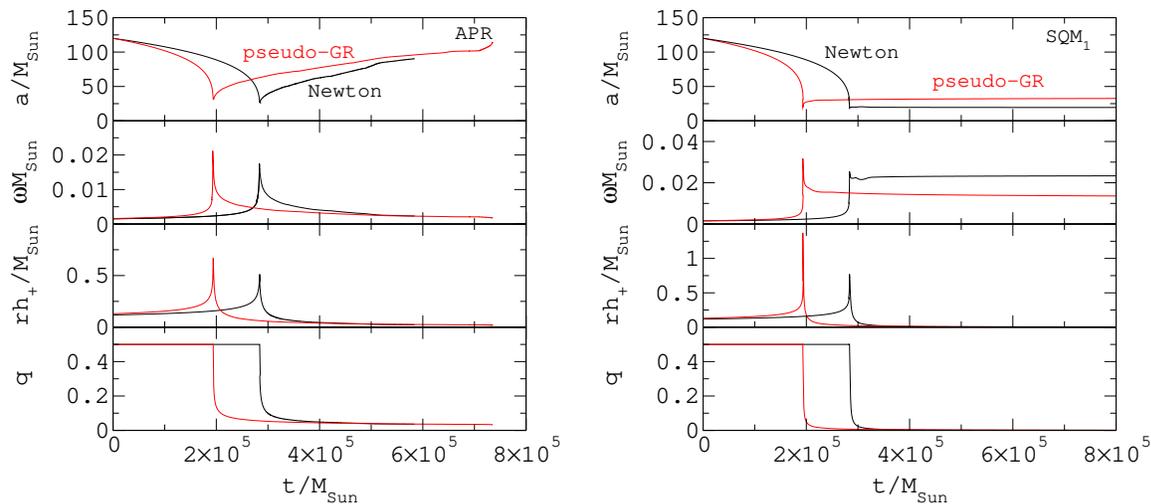}
\end{center}
\caption{\label{evol}Evolution of mergers of a black hole and a normal
star (left panel) and a black hole and an SQM star (right panel). In
all cases, the initial values $M_1=1.33~{\rm M}_\odot$ and $M_2
=2.67~{\rm M}_\odot$ were employed. The EOS symbols are as in
Ref. \cite{LP01}.  From top to bottom: separation distance $a$,
orbital angular frequency $\omega$, waveform
amplitude $h_+$, and mass ratio $q$. }
\end{figure}

Figure~\ref{evol} compares the mergers of normal and self-bound stars
with a BH. Stable mass transfer ensues at the ``kinks'' visible in the
curves for the orbital separation $a$, the orbital frequency $\omega$
and the scalar gravitational polarization amplitude $h_+$. 
The normal star case is shown in the left panel; the self-bound case
is shown in the right panel.  Both Newtonian and pseudo-general
relativistic potential cases are illustrated.  In all cases, 
$\omega(t)$, 
$h_+(t)$ and $q(t)=M_1/M_2$ exhibit abrupt
variations at the onset of mass transfer.  Within each of the four
evolutions considered, variations in the EOS do not qualitatively
alter the results.  The major effect of incorporating general
relativistic corrections to the potential is to speed up the
evolution relative to the Newtonian case.  
Mass transfer thus begins earlier in the pseudo-GR
cases.  This result does not depend upon the properties of the neutron
star.  GR corrections also result in a somewhat larger value for the orbital
separation following the onset of mass transfer.

Larger differences are apparent between the normal and self-bound cases.  The
major differences are: 
\begin{itemize}
\item the orbital separation $a(t)$ increases after mass transfer
begins in the normal case, but only slightly increases
and then remains constant in the self-bound case;
\item reflecting the behavior of $a$, the orbital angular frequency
$\omega(t)$ continuously decreases in the normal case, but quickly
achieves a relatively constant value in the self-bound case;
\item the neutron star mass $M_1$ approaches, but does not fall below,
the minimum mass (about 0.09 M$_\odot$) in the normal case, but in the
self-bound case, the mass dwindles to extremely small values.  In this
case, the smallest mass is that of a strange quark nugget,
determined in part by surface and Coulomb effects; and 
\item the gravitational waveform amplitude $h_+(t)$ follows the
behavior of $q$ and $\omega$ in that they rapidly decay in the
self-bound case, but decay more slowly and remain finite until the end
of mass transfer in the normal matter case.

\end{itemize}

Future tasks will involve the evolution of normal and self-bound
star--black hole mergers including the effects of non-conservative
mass transfer, tidal synchronization, the presence of accretion disks,
etc.  The continued evolution of normal neutron stars beyond the
cessation of stable mass transfer must also be evaluated.

\medskip

This work was supported by the US-DOE grant DE-FG02-87ER40317.




\section*{References}

\begin{thebibliography}{99}

\bibitem{LSch} J.M. Lattimer and D.N. Schramm, 
{Astrophys. J. (Letters)\/}, {\bf 192}, L145 (1974); 
{Astrophys. J.\/} {\bf 210}, 549 (1976). 

\bibitem{Alcock} 
C. Alcock and A. Olinto, 
Ann. Rev. Nucl. Part. Sci. {\bf 38}, 161 (1988). 

\bibitem{LP01}
J.M. Lattimer and M. Prakash, 
Astrophys. J. {\bf 550}, 426 (2001).   






\bibitem{Lee01} W. H. Lee, W. Kluzniak, and J. Nix, Acta Astronomica,
{\bf 51}, 331 (2001); W. Kluzniak and W. H. Lee, MNRAS, {\bf 335}, L29
(2002).


\bibitem{Prakash03} M. Prakash and J.M. Lattimer, Jl. of Phys. G:
Nucl. Part. Phys. {\bf 30}, S451 (2003).

\bibitem{Shibata03} M. Shibata, K. Taniguchi and K. Ury{\=u}, Phys. Rev. D. 
{\bf 68}, 084020 (2003). 

 
\bibitem{Blanchet02} L. Blanchet, Phys. Rev. D. {\bf 65}, 124009 (2002).






 








\end{thebibliography}
\end{document}